# Internet Ware Cloud Computing :Challenges

Dr. S Qamar[1], Niranjan Lal[2], Mrityunjay Singh[3]

[1]Department of Computer Science, CAS, King Saud University,Riyadh, Saudi Arabia

[2,3]Department of Information Technology, SRM University-NCR Campus, Ghaziabad ,India

drsqamar@rediffmail.com, niranjan_verma51@yahoo.com, mrityunjay.cse045@gmail.com.

*Abstract-* **After decades of engineering development and infrastructural investment, Internet connections have become commodity product in many countries, and Internet scale "cloud computing" has started to compete with traditional software business through its technological advantages and economy of scale. Cloud computing is a promising enabling technology of Internet ware Cloud Computing is termed as the next big thing in the modern corporate world. Apart from the present day software and technologies, cloud computing will have a growing impact on enterprise IT and business activities in many large organizations. This paper provides an insight to cloud computing, its impacts and discusses various issues that business organizations face while implementing cloud computing. Further, it recommends various strategies that organizations need to adopt while migrating to cloud computing.The purpose of this paper is to develop an understanding of cloud computing in the modern world and its impact on organizations and businesses. Initially the paper provides a brief description of the cloud computing model introduction and its purposes. Further it discusses various technical and non-technical issues that need to be overcome in order for the benefits of cloud computing to be realized in corporate businesses and organizations. It then provides various recommendations and strategies that businesses need to work on before stepping into new technologies.**

## 1. INTRODUCTION

Everyone has an opinion on what is cloud computing. It can be the ability to rent a server or a thousand servers and run a geophysical modeling application on the most powerful systems available anywhere. It can be the ability to rent a virtual server, load software on it, turn it on and off at will, or clone it ten times to meet a sudden workload demand. It can be storing and securing immense amounts of data that is accessible only by authorized applications and users [13].

Cloud Computing is a broad concept of using the internet to allow people to access technology-enabled services. It is named after the cloud representation of the Internet on a network diagram. Cloud computing is the reincarnation of Centralized data processing and storage as paralleled by the mainframe. A mainframe could be a large computer used by large organizations for bulk data processing. In a broader context, cloud computing is a large network of computers used by large organizations to provide services to smaller ones and individuals. Cloud computing is sometimes also termed as Grid computing or Network computing. Cloud computing is a resource delivery and usage model. It means getting resource via network "on-demand" and "at scale" in a multi-tenant environment. The network of providing resource is called "Cloud". 'What goes on in the cloud manages multiple infrastructures across multiple organizations consisting of frameworks providing mechanisms for self-healing, self-monitoring and automatic reconfiguration', states Kevin Hartig [3].

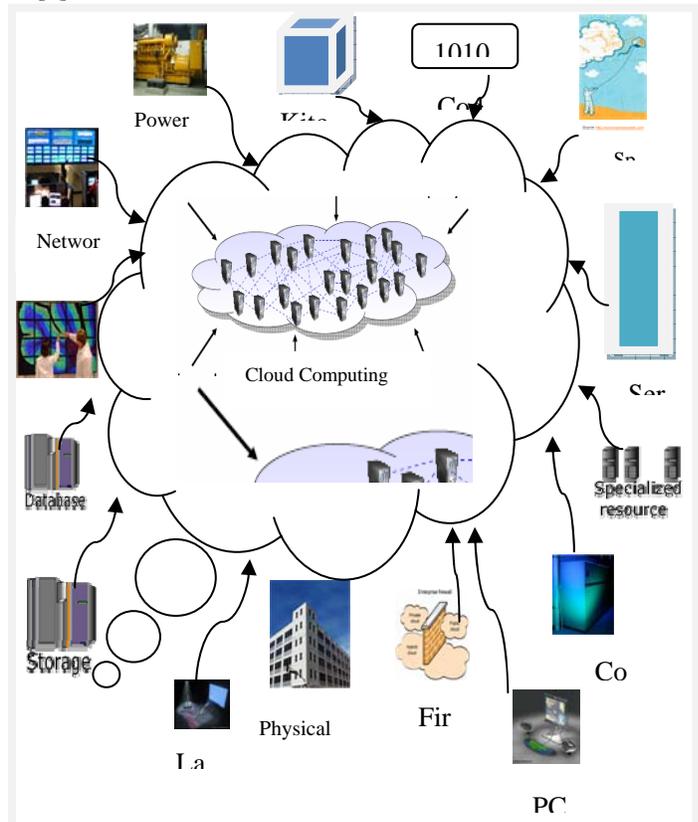

Figure 1. Cloud Computing

The cloud being a virtualization of resources manages itself. Although there are people who maintain and keep track of the hardware, operation systems and networking in a proper order, but from the user's or application developer's perspective, only the cloud is referenced. Cloud computing is the third revolution of IT industry, following the Personal computer revolution and the Internet revolution. The Cloud computing matters to us as cloud computing and web based applications are the future of computing in which all of us will interact. In our daily life, we come across a number of vendors providing cloud computing services such as Gmail, Google, Yahoo, MSN, etc.[4] Among web based office applications and online photo and document sharing include flickr and Zoho. Large scale web based storage and computing power applications to suit our needs include services like Google App Engine and Amazon Web services.In a world where almost anyone and anything can connect to the Internet, the exponential increase in the volume of information and connected devices creates a dilemma: IT complexity increases as does the demand for simplicity. Organizations are facing accelerating business change, global and domestic competitive pressure and social responsibility demands. They are striving to reach their full potential by rapidly implementing innovative business models while simultaneously lowering the IT barriers to driving innovation and change. These challenges call for a more dynamic computing model that enables rapid innovation for applications, services and service delivery. Cloud computing can be one element of such a model. The underlying technologies associated with cloud computing can also be a part of an innovative approach focusing on creation of a more dynamic enterprise, as applications and the services they support are no longer locked to a fixed, underlying infrastructure and can adjust quickly to change. These views were expressed in an official report [14].

By cloud computing, we have the ability to scale to meet changing user demands quickly, usually within minutes. Cloud computing is environmental friendly, task oriented and virtually requires no maintenance. Data is usually not lost in the event of a disaster. It is easy to build one's own web based applications that run in the cloud. In cloud computing, one has the benefit of separation of application code from physical resources and one can use external assets to handle peak loads.

Like any other revolution, cloud computing is the result of the technological process and business model transition. The major driving factors include [3]

The virtualization tech and market's fast development

The hardware's fast development, like CPU and network devices

The wide band's network fast development

The fast increase of corporate IT infrastructure requirement

The fast change and time-to-market requirement of Internet applications

Economy crisis forcing companies to cut cost

## II. Issues to be overcome

Cloud computing is easiest to adopt when there is a considerably flexible approach to phasing it in and relating it to other applications. The biggest challenge in cloud computing may be the fact that there is no standard or single architectural method. Therefore, it is best to view cloud architectures as a set of approaches, each with its own examples and capabilities. Describes below are some of the most common hurdles that need to be overcome in an organization:

### 2.1 Technical Issues

*A. Security*

One of the significant technical hurdles that need to be overcome in order for cloud computing benefits to be realized is security. Reliability and security concerns in an organization might need mitigation and applications might need to be re-architected [7]. Business perceptions of increased IT flexibility and effectiveness will have to be properly managed. In most of the cases, network security solutions are not properly architected to keep up with the movement required for cloud to deliver cost effective solutions. With their businesses' information and critical IT resources outside the firewall, customers worry about their vulnerability to attack.

B. *Technical Hardware & Software Expertise*

Users need equipment and resources to customize cloud computing services more relevant and more tailored to the needs of their businesses. Proper man-power is needed to develop the applications to suit a business's needs. The availability of the physical hardware and software components need to be ensured for realizing the benefits of cloud computing. According to Dion Hinchcliffe [4,5], wider technical fluency and expertise in the selected cloud computing platforms, which tend to emphasize technologies such as Open Source or newer web-style programming languages and application models will have to be achieved.

C. Non-Technical Issues

Apart from the technical issues, there are several non-technical issues which require equal attention and need to be resolved. Some of the significant non-technical hurdles to the adoption of cloud computing services by large enterprises are financial, operational and organizational issues [7].

*D. Financial Issues*

According to McKinsey & Co.'s new report [7]), "Clearing the Air on Cloud Computing", cloud computing can cost twice as much as in-house data centers. This poses a problem for large enterprises, but actually works to the advantage of small and midsize companies and businesses.

McKinsey further states, "Cloud offerings currently are most attractive for small and medium-sized enterprises…and most customers of clouds are small businesses." The reason behind this is that smaller companies don't have the option of developing themselves into giant data centers. Greenberg notes, "Few if any major corporations are looking to replace their data centers with a cloud…the 'server-less company' are one that's only feasible for startups and SMBs." Cost variability is an important aspect of cloud computing. When one considers cost transparency, scalability and cost variability, a new challenge and opportunity for organizations arises.

### *2.2 Operational & Organizational Issues*

Organizations need to define standards and workflow for authorizations. A strategy for the consumption and management of cloud services, including how the organization will deal with semantic management, security and transactions need to be created. One should evaluate cloud providers using similar validation patterns as one does with new and existing data center resources. According to Vinita Gupta [2], before deciding to switch over to cloud computing, one should fully understand the concept and implications of cloud computing as to whether maintaining an IT investment in-house or buying it as a service. The organization has to look at the overall return on investments as they cannot simply rip off and replace an existing infrastructure. The managers have to look at the short-term costs as well as the long-term gains. Service levels offered by different vendors need to be analyzed in terms of uptime, response time and performance. Finally, a proof of concept should be created which can do a few things including getting an organization through the initial learning process and providing proof points as to the feasibility of leveraging cloud computing resources. some Internal Issues that a business might face While switching to newer technologies, an organization could face many internal issues. Some of them are explained as follows:

A. *Distributed business levels*

The distributed business and the level of consistently reliable computer networks in an organization can pose a challenge towards switching from traditional infrastructure to cloud computing. The case for an organization to go in for cloud computing is similar to a decision to own or rent a house. An organization which has spent a good amount of cash on its own storage and security systems will have a tough time taking the decision to migrate to a dedicated environment.

*B. Complexity of applications*

The complexity of the applications and the technology infrastructure is dependent on how the organization has adopted IT. If this has evolved from the deployment of technologies over a period of time, then the complexity level will certainly be high and in such a case, transformation to cloud computing would be difficult. Not everything comes under cloud computing as each organization has its own specific requirements suited to their needs whether on functionalities, performance, or maybe even security and privacy needs that could be unique to an organization, and may not be supported through the public cloud[2]. It is relatively very difficult to adapt to cloud computing in an organization where highly customized applications or home grown applications are used. The availability of a robust network and information security is also a challenge.

*C. Cost*

Cost of process change is another hurdle in the transformation. Conventional IT organizations will have to engage with internal customers as well as IT service providers at a different plane. Most importantly, the culture and mindset will have to change.

### *III. Security & Reliability Issues*

Bryan Gardiner [1] expresses that for large corporate organizations, a number of concerns regarding the adoption of cloud computing arise. In most of the cases, they are worried about lost or stolen data. Hard-liners see the very concept of the cloud as a deeply unreliable security nightmare. In a survey conducted by a research firm IDC, almost 75 percent of I.T. executives reported that security was their primary concern, followed by performance and reliability [1], i.e., how enterprise data is safeguarded in a shared third-party environment. The pace of the future uptake is heavily dependent on how soon these issues are resolve, and when cloud providers will be able to obtain official certifications of their security practices from independent third parties.

In spite of the work-in-progress, a large number of other research issues remain and part of the purpose of this paper is to help define a longer-term research agenda for the researchers. Key issues include:

How do consumers know what services are available and how do they evaluate them?

How do consumers express their requirements?

How are services composed?

How are services tested?

What is the appropriate, high integrity, service delivery infrastructure?

How must consumers' data be held to enable portability between different service suppliers?

What standards can be used or must be defined to enable portability of service?

What will be the impact of branded services and marketing activities (high quality v low price)?

How can organizations benefit from rapidly changing services and how will they manage the interface with business processes?

How will individuals perceive and manage rapidly changing systems? What is the limit to the speed of change?

What payment and reward structures will be necessary to encourage SME service suppliers?

What will be the new industry models and supply chain arrangements?

How can we evaluate the research outcomes?

How to plan towards the new technologies?

Cloud computing is inevitable and it is a force that organizations and businesses need to quickly come in terms with. As the economic and social motivation for cloud computing is high, businesses which are heavily computer resource dependent need to take cautionary measures and the right decisions at the right time to avoid ending up with unproductive solutions while migrating to new technologies. According to Davis Robbins [10], an organization should always make sure that they know what they are paying for and should pay careful attention to the following issues:

Service levels

Privacy matters

Compliances

Data ownership

Data mobility

A number of cloud computing vendors may be hesitant to commit to the consistency of performance regarding an application or transaction. One has to understand the service levels they expect regarding data protection and speed of data recovery.

In large corporate organizations, privacy matters a lot. Someone else hosting and serving the organizations' data could be approached by someone else from within or outside the cloud without one's knowledge or approval. All the regulations applying to the business must properly be reviewed. Cloud services and vendors must meet the same level of compliance for data stored in the cloud. One has to make sure that if a cloud relationship is terminated, would the data be shared between the cloud services and the organization. If it is returned back, which format it will be in. Development and test activities must be carried out prior to switching completely to cloud services. It will allow the organization to reduce capital spending and related data center costs while increasing speed and agility. In addition to this, one can also evolve their internal infrastructure towards a more cloud-like model. One needs to identify which services can reside in the cloud and which should be present internally within the business. Systems and services core to the business should be determined and a sourcing strategy to achieve the low cost, scalability and flexibility should be determined. This should include all the necessary protections such as data ownership, mobility and compliance.

Businesses must keep costs down to stay competitive while at the same time investing in new ideas that will provide compelling and attractive new products and services to their customers.

Businesses should be able to compare their traditional computer cost system to the utility pricing model common in the cloud computing business. Other hidden costs such as management, governance and transition costs including the hiring of new staff must also be considered. The amount of time required to recoup the investment in a transition to cloud computing need to be analyzed. Company executives need to discuss important issues while bringing cloud computing into the picture such as the effect of service-oriented architecture strategy, impaction of disaster recovery plans, policies regarding backups and legally mandated data archives, risk profile for using cloud computing services and mitigation strategies.

### IV. Switch over to new technologies

Switching to newer technologies such as cloud computing would be best when the processes, applications, and data are largely independent. When the points of integration in a business are well defined, embracing cloud services is effective. In an organization where a lower level of security will work just fine and the core internal enterprise architecture is healthy, conditions are favorable for the organization to switch to newer technologies. A business which requires Web as the desired platform to serve its customers and wants to cut cost while benefiting from the new applications, the business can achieve the best competitive advantage in the market.

There for to compete effectively in today's world, executives need every edge they can get, from low cost to speed and employee productivity. By tapping into the right cloud capabilities, companies can quickly enter new markets and launch new products or services in existing markets. When demand grows, they can quickly scale up, and when opportunities dry up, they can just as quickly scale down with minimum waste of time and capital. By using cloud-based solutions such as crowd-sourcing, companies can open up innovation to more employees, customers and their partners.

### V. Conclusion and Future work

Cloud computing is a fascinating realm, that makes it easier to deploy software and increase productivity. However, there are some technical and non-technical realities that make security somewhat difficult to deliver in a cloud. The cloud presents a number of new challenges in data security, privacy control, compliance, application integration and service quality. It can be expected that over the few years, these problems will be addressed. There for to be successful, companies should take small incremental steps towards this new environment so they can reap benefits for applicable business situations and learn to deal with the associated risks. In general, cloud computing will act as an accelerator for enterprises, enabling them to innovate and compete more effectively. Under the current economic conditions, executives need to rethink their strategies dealing with cost-effective solutions. They need to use the cloud services for the right jobs they require. Today's infrastructure clouds such as Amazon EC2 offer a relatively inexpensive and flexible alternative to buying in-house hardware. They are also beneficial for computation-intensive jobs, such as data cleansing, data mining, risk modeling, optimization and simulation. Businesses and enterprises should now take steps to experiment, learn and reap some immediate business benefits by implementing cloud computing in their organizations. Unless they seriously consider making the cloud a part of their strategy, they would find themselves disadvantaged when competing in today's increasingly multi-polar marketplace. The number of mobile applications is growing sharply. However, limited processing power, battery life and data storage of mobile phone must limit the growth of application software for mobile industry. The adoption of cloud computing should be make mobile applications more sophisticated and make them to available for broader audience of subscribers.